# Transition of laser-induced terahertz spin currents from torque- to conduction-electron-mediated transport

Pilar Jiménez-Cavero[1,2,3,4,*], Oliver Gueckstock[1,2,*], Lukáš Nádvorník[1,2,5,†], Irene Lucas[3,4] Tom S. Seifert[1,2], Martin Wolf[2], Reza Rouzegar[1,2], Piet W. Brouwer[1], Sven Becker[6], Gerhard Jakob[6], Mathias Kläui[6], Chenyang Guo[7,8], Caihua Wan[7], Xiufeng Han[7,8], Zuanming Jin[9,10], Hui Zhao[11], Di Wu[11], Luis Morellón[3,4], Tobias Kampfrath[1,2,†]

1. Department of Physics, Freie Universität Berlin, Arnimallee 14, 14195 Berlin, Germany
2. Department of Physical Chemistry, Fritz Haber Institute of the Max Planck Society, Faradayweg 4-6, 14195 Berlin, Germany
3. Instituto de Nanociencia y Materiales de Aragón (INMA), Universidad de Zaragoza-CSIC, Mariano Esquillor, Edificio I+D, 50018 Zaragoza, Spain
4. Departamento Física de la Materia Condensada, Universidad de Zaragoza, Pedro Cerbuna 12, 50009 Zaragoza, Spain
5. Faculty of Mathematics and Physics, Charles University, Ke Karlovu 3, 12116 Prague, Czech Republic
6. Institut für Physik, Johannes Gutenberg-Universität Mainz, 55128 Mainz, Germany
7. Beijing National Laboratory for Condensed Matter Physics, Institute of Physics, University of Chinese Academy of Sciences, Chinese Academy of Sciences, Beijing 100190, China
8. Center of Materials Science and Optoelectronics Engineering, University of Chinese Academy of Sciences, Beijing 100049, China
9. Shanghai Key Lab of Modern Optical Systems, University of Shanghai for Science and Technology, Shanghai 200093, China
10. Department of Physics, Shanghai University, Shanghai 200444, China
11. Department of Physics and National Laboratory of Solid State Microstructures, Nanjing University, 210093, China

[*] contributed equally to this work
[†] E-mail: nadvornik@karlov.mff.cuni.cz, tobias.kampfrath@fu-berlin.de

Spin transport is crucial for future spintronic devices operating at bandwidths up to the terahertz (THz) range. In F|N thin-film stacks made of a ferro/ferrimagnetic layer F and a normal-metal layer N, spin transport is mediated by (1) spin-polarized conduction electrons and/or (2) torque between electron spins. To identify a cross-over from (1) to (2), we study laser-driven spin currents in F|Pt stacks where F consists of model materials with different degrees of electrical conductivity. For the magnetic insulators YIG, GIG and $\gamma$-$Fe_2O_3$, identical dynamics is observed. It arises from the THz interfacial spin Seebeck effect (SSE), is fully determined by the relaxation of the electrons in the metal layer and provides an estimate of the spin-mixing conductance of the GIG/Pt interface. Remarkably, in the half-metallic ferrimagnet $Fe_3O_4$ (magnetite), our measurements reveal two spin-current components with opposite direction. The slower, positive component exhibits SSE dynamics and is assigned to torque-type magnon excitation of the A- and B-spin sublattices of $Fe_3O_4$. The faster, negative component arises from the pyro-spintronic effect and can consistently be assigned to ultrafast demagnetization of e-sublattice minority-spin hopping electrons. This observation supports the magneto-electronic model of $Fe_3O_4$. In general, our results provide a new route to the contact-free separation of torque- and conduction-electron-mediated spin currents.

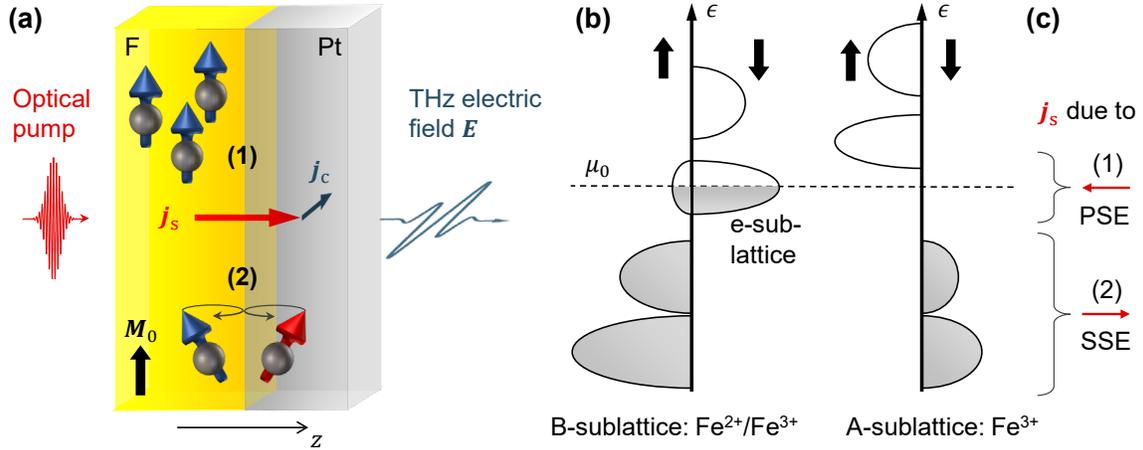

**FIG. 1.** (a) Schematic of photoinduced spin transport in an F|Pt stack, where Pt is platinum, and F is a magnetic layer with equilibrium magnetization $M_0$. An ultrashort laser pulse excites the sample and induces an ultrafast spin current with density $j_s$ from F to Pt along the $z$ axis. In the Pt layer, $j_s$ is converted into a transverse charge current with density $j_c$ that gives rise to the emission of a terahertz (THz) electromagnetic pulse. Schematics (1) and (2) indicate spin transfer across the F/Pt interface that is mediated by (1) spin-polarized conduction electrons and (2) spin torque, the latter coupling to magnons in F. Both (1) and (2) can be driven by gradients of temperature and spin accumulation. (b) Simplified schematic of the single-electron density of states vs electron energy $\epsilon$ of the tetrahedral A- and octahedral B-sites of the ferrimagnetic half-metal $Fe_3O_4$. The DC conductivity predominantly arises from the B-site minority-type hopping electrons of the e-sublattice. (c) In our interpretation, optical excitation of the $Fe_3O_4$|Pt stack triggers spin transfer through both the spin Seebeck effect (SSE) and pyro-spintronic effect (PSE). While the SSE current is mediated by torque between Pt and $Fe_3O_4$ electron spins far below the Fermi level $\mu_0$ [(2) in panel (a)], the PSE transiently increases the chemical potential of the B-site minority-spin electrons around $\mu_0$ and, thus, induces a conduction-electron spin current [(1) in panel (a)].

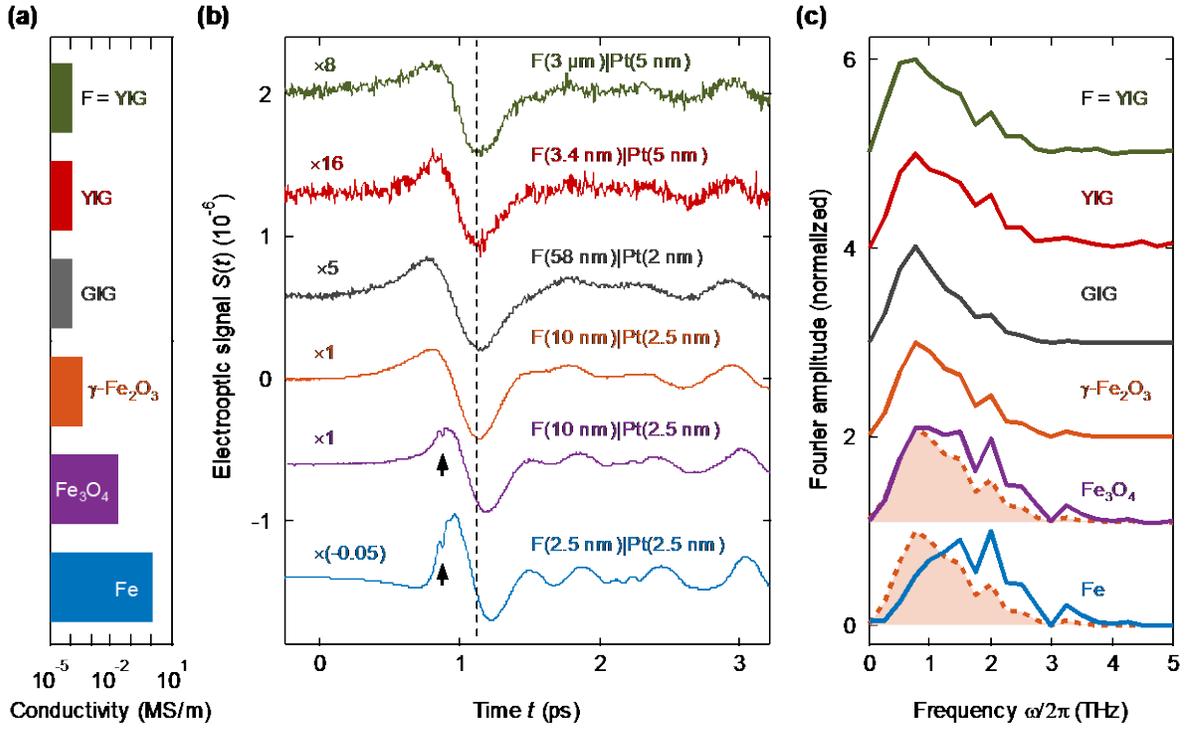

**FIG. 2.** THz emission from F|Pt bilayers as a function of F-layer conductivity. (a) Electrical conductivities of the studied F materials on a logarithmic scale. (b) Electro-optic signals of THz pulses emitted from various F|Pt bilayers with F=YIG (thick and thin), GIG, $\gamma$-Fe$_2$O$_3$, Fe$_3$O$_4$ and Fe. Note the different amplitude scaling's. The time-axis origin is the same for all signals and was determined by the signal from Fe|Pt reference stacks (Fig. S1). The dashed vertical line marks the minimum signal for the insulating F materials YIG, GIG and $\gamma$-Fe$_2$O$_3$, and the two black arrows label a sharp feature in the traces for F=Fe$_3$O$_4$ and Fe. (c) Fourier amplitude spectra of the signals of panel (b) (normalized to peak height 1). Dashed lines show two duplicates of the spectrum of $\gamma$-Fe$_2$O$_3$|Pt. Curves in (b) and (c) are vertically offset for clarity.

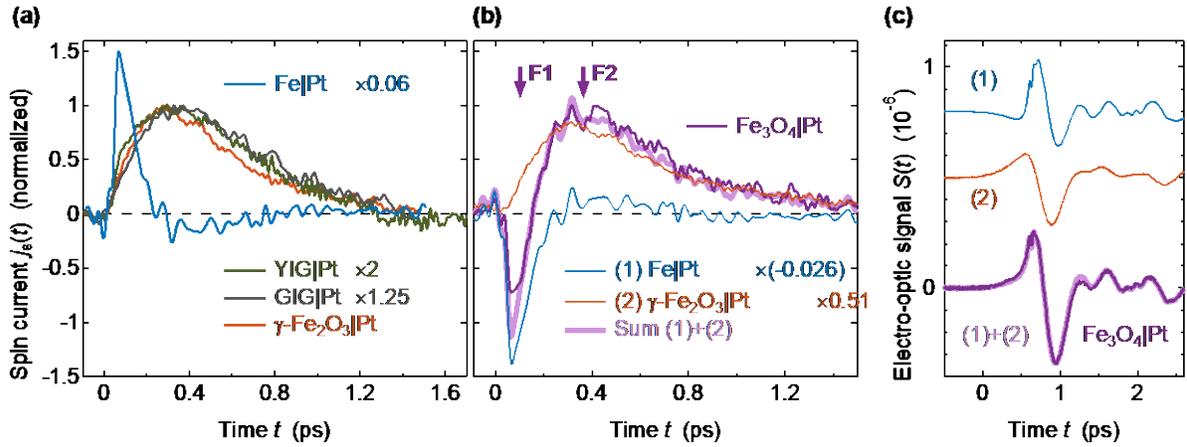

**FIG. 3.** Ultrafast photoinduced spin transport in F|Pt stacks. (a) Curves show the spin current density $j_s(t)$ in magnetic-insulator|Pt and Fe|Pt stacks, i.e., YIG(3 µm)|Pt(5 nm), GIG(58 nm)|Pt(2 nm), $\gamma$-Fe$_2$O$_3$(10 nm)|Pt(2.5 nm) and Fe(2.5 nm)|Pt(2.5 nm), as extracted from the THz emission signals of Fig. 2b. Each signal is normalized by the pump-excitation density inside the Pt layer and by the indicated factor. (b) Spin current $j_s(t)$ in Fe$_3$O$_4$(10 nm)|Pt(2.5 nm) along with scaled spin currents in $\gamma$-Fe$_2$O$_3$|Pt and Fe|Pt. The violet arrows F1 and F2 mark characteristic features of the curves. Note that $j_s(t)$ in Fe$_3$O$_4$|Pt can be well described as a linear combination of the other two spin currents (light-violet curve). (c) Same as in panel (b), but for the THz-emission raw signals.

## I. INTRODUCTION

**Spin currents**

Control over spin currents is a cornerstone of spintronic technologies [1]. New functionalities in such diverse fields as energy conversion and information technologies are envisaged to benefit from the generation, processing and detection of spin currents [2-5]. An important goal is to push the bandwidth of spintronic operations to the terahertz (THz) frequency range, corresponding to ultrafast time scales [1].

A model system for the investigation of the transport of spin angular momentum is the F|N thin-film stack of Fig. 1(a), where spin can be transmitted from a ferro- or ferrimagnetic layer F to an adjacent non-ferro/ferrimagnetic metal layer N. The spin current in F is mediated not only by (1) spin-polarized conduction electrons, which typically dominate spin transfer in metals, but also by (2) magnons, i.e., torque between coupled spins [6,7], which is the main transport channel in insulators. Accordingly, spin transfer across an F/N interface can be mediated by (1) spin-polarized conduction electrons traversing the interface [see (1) in Fig. 1(a)] and by (2) spin torque between adjacent F and N regions [(2) in Fig. 1(a)]. As mechanism (2) results in the excitation of magnons in F [8], it can be considered as magnonic spin transfer.

In general, to drive an incoherent spin current of density $j_s$ from F to N, a difference in temperature or spin chemical potential (also known as spin accumulation or spin voltage) between the two layers is required [9,10]. For example, for a temperature gradient between F and N, the resulting spin current arises from the interfacial spin-dependent Seebeck effect (SDSE) [11] for channel (1) or the interfacial spin Seebeck effect (SSE) [12-16] for channel (2).

In any case, the spin flow from F to N can be detected by conversion of the longitudinal $j_s$ into a transverse charge current with density $j_c$ [Fig. 1(a)] and measurement of the resulting voltage. For this purpose, N materials with sufficiently large inverse spin Hall effect (ISHE), for instance Pt, are well suited.

**THz spin transport**

A powerful and ultrafast approach to deposit excess energy in F|N stacks is optical excitation by femtosecond laser pulses [Fig. 1(a)]. Measurement of the ultrafast transverse charge current $j_c$ as a function of time $t$ allows one to resolve elementary relaxation processes such as electron thermalization [8] and electron-spin and electron-phonon equilibration [10], but also delivers insights into spin-to-charge-current conversion [17-24].

For an insulating and pump-transparent F, temperature gradients between F and N (i.e., the SSE) were found to be the dominant driving force of the ultrafast $j_s$ [8,25]. For metallic F, in contrast, such temperature differences (i.e., the SDSE) were concluded to make a minor contribution. Instead, spin-voltage-gradients were suggested and identified as the relevant driving force of spin transport on sub-picosecond time scales in metals [10,26-30]. More precisely, dynamic heating of F leads to a transient spin accumulation or spin voltage, which quantifies the excess of spin angular momentum in F. This phenomenon, which may be termed pyro-spintronic effect (PSE), induces a spin current across the F/N interface [10,28].

There are important open questions regarding the role of THz SSE and PSE. First, the SSE was so far only observed in F|Pt stacks with F made of yttrium iron garnet (YIG). According to the microscopic model of Ref. [8], the spin-current dynamics should be fully determined by the relaxation dynamics of the Pt electrons, independent of the insulating F-layer material. This quite universal implication remains to be shown.

Second, with increasing electrical conductivity of the F material, a transition from ultrafast SSE to PSE should occur, which was not yet observed. At the cross-over point, both spin transport channels

(1) and (2) may be operative [Fig. 1(a)], and disentangling them is crucial to maximize the overall generation efficiency of spin currents. The experimental separation of conduction-electron- and magnon-carried incoherent spin transport is challenging under quasistatic conditions [8,31,32]. However, on femtosecond time scales, SSE and PSE dominate and exhibit different temporal dynamics, as indicated by previous works [8,10]. Thus, a separation of the two effects might be possible.

**This work**

In this work, we study ultrafast photogenerated spin currents in F|Pt bilayers as a function of various magnetic model F-materials with different degrees of electric conductivity: The ferrimagnetic insulators maghemite ($\gamma$-Fe$_2$O$_3$), gadolinium iron garnet (Gd$_3$Fe$_5$O$_{12}$, GIG) and YIG (with a thickness ranging over three orders of magnitude), the ferrimagnetic half-metal magnetite (Fe$_3$O$_4$), and, for referencing purposes, the ferromagnetic metal iron (Fe).

Our study reveals that the ultrafast dynamics of the SSE is independent of the choice of the magnetic insulator (YIG, GIG, $\gamma$-Fe$_2$O$_3$), its thickness (3.4 nm-3 µm) and growth method. Remarkably, in the half-metallic ferrimagnet Fe$_3$O$_4$, we observe simultaneous signatures of both SSE and PSE, whose ultrafast spin currents counteract each other. The PSE current is much smaller and of opposite sign compared to Fe. We assign the PSE current in Fe$_3$O$_4$ to the minority hopping electrons (Fig. 1b).

## II. EXPERIMENTAL DETAILS

**THz emission setup**

To launch an ultrafast spin current, the sample under study is excited with near-infrared femtosecond laser pulses (central wavelength of 800 nm, duration of 10 fs, energy of 1 nJ, repetition rate of 80 MHz) from a Ti:sapphire laser oscillator [see Fig. 1(a)]. Part of the energy of the incident pump pulse is instantaneously deposited in the electronic system of the Pt layer and, if metallic, of F. Any induced spin current $j_s(t)$ flowing across the F/Pt interface is partially converted into a transverse charge current $j_c(t)$ in Pt through the ISHE, thereby resulting in the emission of electromagnetic pulses with frequencies extending into the THz range [Fig. 1(a)] [8,17,19-24,33,34]. We detect the transient THz electric field by electro-optic sampling in a 1 mm thick ZnTe(110) crystal, resulting in the electrooptic signal $S(t, \boldsymbol{M}_0)$ [35-37].

During the experiments, the in-plane equilibrium magnetization $\boldsymbol{M}_0$ of the sample is saturated by an external magnetic field with (magnitude $\approx$100 mT). We measure signals for two opposite orientations $\pm \boldsymbol{M}_0$. Because we are only interested in effects odd in the magnetization $\boldsymbol{M}_0$, we focus on the antisymmetric signal

$$S(t) = \frac{S(t, +\boldsymbol{M}_0) - S(t, -\boldsymbol{M}_0)}{2}. \tag{1}$$

All data are acquired at room temperature in air if not mentioned otherwise.

**Material choice**

For the F material in our F|Pt stacks, we choose common and spintronically relevant two-lattice ferrimagnets with increasing degree of electric conductivity: (i) insulating YIG (thickness 3.4 nm-3 µm), (ii) insulating Gd$_3$Fe$_5$O$_{12}$ (58 nm), (iii) insulating $\gamma$-Fe$_2$O$_3$ (10 nm) and (iv) the half-metal Fe$_3$O$_4$ (10 nm) [38]. For referencing, (v) the ferromagnetic metal Fe (2.5 nm) is chosen. As N material, we choose Pt for all samples due to its large spin Hall angle [39]. The approximate F-material conductivities [40-43] are summarized in Fig. 2(a).

The insulating F materials transfer spin angular momentum by torque [Fig. 1(a), (1)], whereas for the metal Fe, the spin current is expected to be carried predominantly by conduction electrons [Fig. 1(a), (2)] [10].

In this respect, $Fe_3O_4$ is special because it exhibits both localized and mobile electrons with magnetically ordered spins. More precisely, the ferrimagnet $Fe_3O_4$ is a half-metal, since its conductivity is dominated by hopping-type transport of minority electrons. $Fe_3O_4$ possesses two sublattices A and B of, respectively, localized $Fe^{2+}/Fe^{3+}$ and $Fe^{3+}$ spins, which couple antiferromagnetically [38]. In the so-called magneto-electric model, the spins of the hopping electrons are aligned predominantly antiparallel to $\boldsymbol{M}_0$ due to exchange interaction with A and B and, thus, form the e-sublattice [44-47].

A highly simplified schematic of the electronic structure of $Fe_3O_4$ is displayed by the spin- and site-resolved single-electron density of states in Fig. 1(b) [48,49]. The majority (spin-up) electrons exhibit an electronic band gap with a calculated magnitude of 1.7 eV [50], while the presence of minority (spin-down) hopping electrons at the Fermi level $\mu_0$ [44] makes magnetite a half-metal. The measured spin polarization at $\mu_0$ amounts to -72% in $Fe_3O_4$(001), indicating a nonvanishing number of majority hopping electrons [Fig. 1(b)] [51].

**Sample details and excitation**

Details on the sample fabrication can be found in the Appendix A. In brief, the YIG films are fabricated by three different techniques (pulsed-laser deposition, sputtering and liquid-phase epitaxy). The Fe|Pt reference sample is obtained by growing an Fe layer on top of half the F|Pt region for most of the samples [Fig. S1]. The THz emission signal from the resulting F|Pt|Fe regions is dominated by Pt|Fe and equals the reversed signal from an Fe|Pt layer [8]. By means of the Fe|Pt reference signals, the time axes of the THz signals from all samples can be synchronized with an accuracy better than 10 fs.

The pump electric field is approximately constant along $z$ [Fig. 1(a)] throughout the thin-film stack of our samples. Therefore, the locally absorbed pump-pulse energy is only proportional to $\text{Im}(n^2)$, where $n$ is the complex-valued refractive index of the material at the pump wavelength (800 nm). While the Pt and Fe layers are strongly absorbing [$\text{Im}(n^2)$ =28 and 30] [52], $Fe_3O_4$ is weakly absorbing (2.3) [53], and YIG, GIG and $\gamma$-$Fe_2O_3$ are largely transparent to the pump beam [$\text{Im}(n^2) \lesssim 1.5$] [54,55].

## III. RESULTS AND DISCUSSION

**Terahertz emission signals**

Figure 2(b) shows electro-optic signals $S(t)$ [Eq. (1)] of THz pulses emitted by the Fe|Pt, $\gamma$-$Fe_2O_3$|Pt, $Fe_3O_4$|Pt, GIG|Pt and the thinnest as well as the thickest YIG|Pt samples. THz signals from all other YIG samples can be found in Fig. S2(a). Measurements of YIG(3 µm)|Pt(5 nm) [8] and $Fe_3O_4$|Pt confirm that the THz signal increases linearly with the pump power [Fig. S7]. We make two observations in terms of (i) signal shape and (ii) magnitude.

(i) The waveforms from all samples with YIG, GIG and $\gamma$-$Fe_2O_3$ exhibit very similar dynamics [Fig. 2(b) and Fig. S3]. In contrast, the signal for $Fe_3O_4$ features a steeper initial rise, a sharp notch right before the first maximum (see black arrow) and a subsequent smaller peak. The global minimum is shifted to later times, as indicated by the dashed vertical line. This trend is even more enhanced for Fe|Pt. These observations indicate that different processes occur in the samples as the F-material conductivity increases [Fig. 2(a)] [38,40,41,56,57].

(ii) While the signals from all YIG-based samples have similar strengths [Fig. S2(a)], the signals from the $\gamma$-$Fe_2O_3$ and $Fe_3O_4$ samples are nearly one order of magnitude larger. The signal from Fe|Pt is even more than two orders of magnitude larger than from YIG|Pt.

By Fourier transformation of the time-domain waveforms $S(t)$ [Fig. 2(b)], the normalized amplitude $|S(\omega)|$ as a function of frequency $\omega/2\pi$ is obtained [Fig. 2(c)]. As expected from the time-domain data [Fig. 2(b)], the THz signal of the YIG, GIG and $\gamma$-Fe$_2$O$_3$ samples have approximately the same amplitude spectrum. For Fe$_3$O$_4$, however, a slightly blue-shifted spectrum with an increased bandwidth is found. This trend is more pronounced for the Fe|Pt spectrum.

**Spin current for insulating F materials**

As detailed in Appendix B, we retrieve the spin current dynamics from the measured THz signal waveforms. Figure 3(a) displays the resulting spin current density $j_s(t)$ vs time $t$ in $\gamma$-Fe$_2$O$_3$(10 nm)|Pt(2.5 nm), GIG(58 nm)|Pt(2 nm) and the YIG(3 µm)|Pt(5 nm) samples. We observe that (i) the $j_s(t)$ in GIG|Pt, $\gamma$-Fe$_2$O$_3$|Pt and all YIG|Pt samples exhibit very similar temporal dynamics. (ii) The overall amplitude of the spin current in $\gamma$-Fe$_2$O$_3$|Pt is about one order of magnitude larger than for the YIG|Pt samples. Observations (i) and (ii) are fully consistent with the temporal shape and global amplitude of the underlying raw data [see Fig. 2(b)]. They have three important implications.

*SSE dynamics.*--First, it is remarkable that the optically induced spin currents in F|Pt bilayers proceed with the same dynamics, even though the magnetic layer is made of very different insulators (F=YIG, GIG and $\gamma$-Fe$_2$O$_3$) and covers, in the case of YIG, three different growth techniques. Note that in these samples, the pump pulse is to the largest extent absorbed by the Pt layer. Therefore, observation (i) confirms the previous notion [8] that the ultrafast dynamics of the optically induced SSE current are solely determined by the relaxation dynamics of the electrons in the Pt layer.

More precisely, the instantaneous spin current density was predicted to monitor the instantaneous state of the electronic system of N=Pt through [8]

$$j_s(t) = \mathcal{K}\Delta\tilde{T}_e^N(t). \qquad (2)$$

Here, $\mathcal{K}$ is the interfacial spin Seebeck coefficient, and $\Delta\tilde{T}_e^N$ is the pump-induced change in a generalized temperature of the N electrons, which is also defined for nonthermal electron distributions. Importantly, $\Delta\tilde{T}_e^N$ approximately scales with the number of pump-induced electrons above the Fermi level $\mu_0$. Therefore, it is relatively small directly after optical excitation, but subsequently increases by nearly two orders of magnitude owing to carrier multiplication through electron-electron scattering [8]. The rise of $j_s(t)$ on a time scale of 100 fs [Fig. 3(a)], thus, reflects the evolution of the initially nonthermal electron distribution to a Fermi-Dirac distribution. The decay is determined by energy transfer from the electrons to the phonons.

*Impact of YIG thickness.*--Second, finding (i) also implies that the dynamics of the spin current are independent of the YIG thickness, which covers a wide range from 3.4 nm to 3 µm [Fig. S2(b)]. This result supports the notion [8] that the spin current traversing the YIG/Pt interface stems from YIG regions less than a few nanometers away from the YIG/Pt interface. It is easily understood given that magnons in YIG have a maximum group velocity of about 10 nm/ps [58] and that the majority of the ultrafast spin-current dynamics proceed within less than 1 ps [Fig. 3(a)].

*SSE amplitude.*--Third, we observe that the spin current in the $\gamma$-Fe$_2$O$_3$|Pt sample is about 2 times higher than for the YIG|Pt or GIG|Pt sample. To understand how this observation is related to the F/Pt interface, we consider Eq. (2) and note that the SSE coefficient scales according to [8]

$$\mathcal{K} \propto g_r^{\uparrow\downarrow} M_{IF} a^3. \qquad (3)$$

Here, $g_r^{\uparrow\downarrow}$ is the real part of the spin-mixing conductance of the F/Pt interface, $M_{IF}$ is the interfacial saturation magnetization, and $a$ is the lattice constant of F. To obtain the relative magnitude of $g_r^{\uparrow\downarrow}$, we divide the THz peak signal of each YIG, GIG and $\gamma$-Fe$_2$O$_3$ sample by the deposited pump energy density, the THz impedance of the sample, and $M_{IF}a^3$, where bulk magnetization values are assumed for $M_{IF}$ [59-61] (see Appendix B and Table B1).

We infer that $g_\mathrm{r}^{\uparrow\downarrow}$ is very similar in all three materials and has a relative magnitude of 1, 1 and 1.2. Thus, the spin-mixing conductance of the $\gamma$-Fe$_2$O$_3$/Pt and GIG/Pt interfaces approximately equals that of the YIG/Pt interface [62]. We are not aware of any previous $g_\mathrm{r}^{\uparrow\downarrow}$ measurement of GIG/Pt.

**Spin current in Fe|Pt**

The ultrafast pump-induced spin current in the Fe|Pt reference sample is shown in Fig. 3(a) (blue curve). It rises and decays much faster than the SSE-type spin currents in the F|Pt samples with magnetic insulator [Fig. 3(a)].

In a previous work [10], the spin-current dynamics in F|Pt stacks with ferromagnetic metallic F was explained by the PSE: Excitation by the pump pulse leads to a sudden increase of the electron temperature of F as well as of the spin voltage $\Delta\tilde{\mu}_\mathrm{s}$, also called spin accumulation, which quantifies the instantaneous excess of spin density in F. As the system aims to adapt the F magnetization to the excited electronic state, spin angular momentum is transferred from the electrons to the crystal lattice of F and/or to the adjacent Pt layer. Remarkably, temperature gradients between F and Pt (i.e., the SDSE) were concluded to make a minor contribution on sub-picosecond time scales [10], resulting in the simple relationship

$$j_\mathrm{s}(t) \propto \Delta\tilde{\mu}_\mathrm{s}(t). \qquad (4)$$

In the case of Fermi-Dirac distributions, $\Delta\tilde{\mu}_\mathrm{s}$ equals the difference of the chemical potentials of spin-up and spin-down electrons, but the concepts of generalized spin voltage and temperature still apply for non-thermal electron distributions [10].

The transfer of spin angular momentum out of the F electrons into the crystal lattice or the Pt layer leads to a decay of the spin voltage on time scale $\tau_\mathrm{es}$. The dynamics of $j_\mathrm{s}(t)$ is, thus, governed by $\tau_\mathrm{es}$ and the relaxation of the electron excess energy of F, as parameterized by the generalized electron excess temperature $\Delta\tilde{T}_\mathrm{e}^\mathrm{F}$. Quantitatively, the dynamics of $\Delta\tilde{\mu}_\mathrm{s}(t)$ and, thus, $j_\mathrm{s}(t)$ can be described by [10]

$$\Delta\tilde{\mu}_\mathrm{s}(t) \propto \Delta\tilde{T}_\mathrm{e}^\mathrm{F}(t) - \int_0^\infty \frac{\mathrm{d}\tau}{\tau_\mathrm{es}}\, \mathrm{e}^{-\frac{\tau}{\tau_\mathrm{es}}} \Delta\tilde{T}_\mathrm{e}^\mathrm{F}(t-\tau). \qquad (5)$$

Following excitation by the pump [10], $\Delta\tilde{T}_\mathrm{e}^\mathrm{F}$ immediately jumps to a nonzero value. The spin voltage $\Delta\tilde{\mu}_\mathrm{s}(t)$ and $j_\mathrm{s}(t)$ follow without delay, according to the first term of Eq. (5). Due to the subsequent transfer of spin angular momentum out of the F electrons, the spin voltage decays with time constant $\tau_\mathrm{es}$, as forced by the second term of Eq. (5).

As a consequence, the spin current in Fe|Pt rises instantaneously within the time resolution of our experiment (~40 fs) [10], much faster than in, for instance, YIG|Pt [Fig. 3(a)]. Its decay is predominantly determined by electron-spin equilibration on the time scale $\tau_\mathrm{es}$, with a minor correction due to the significantly slower electron-phonon equilibration [10]. To summarize, the very different dynamics of SSE (magnetic-insulator|Pt) and PSE (Fe|Pt) seen in Fig. 3(a) suggest that both effects and, thus, torque- and conduction-electron-mediated spin transport can be separated.

**Spin current in Fe$_3$O$_4$|Pt**

Figure 3(b) displays the spin current $j_\mathrm{s}(t)$ flowing from Fe$_3$O$_4$ to the Pt layer. We observe two features with different dynamics: (F1) A fast and sharp negative dip (see violet arrow F1), followed by (F2) a slower positive feature (arrow F2) that decays with a time constant of 0.3 ps. As Fe$_3$O$_4$ is a half-metal, it is interesting to compare the dynamics in Fe$_3$O$_4$|Pt to those in the two F|Pt stacks with the insulator F=$\gamma$-Fe$_2$O$_3$ and the metal F=Fe [see Fig. 3(b)]. For F=$\gamma$-Fe$_2$O$_3$, the spin current across the F/Pt interface is mediated by spin torque, whereas for F=Fe, it is predominantly carried by spin-polarized electrons.

Note that the fast feature (F1) is comparable to $j_s(t)$ of Fe|Pt (blue curve), whereas the slower feature (F2) resembles the $j_s(t)$ of $\gamma$-Fe$_2$O$_3$|Pt (orange curve). As shown in Fig. 3(b), we are even able to reproduce the $j_s(t)$ of Fe$_3$O$_4$|Pt by a sum of $-0.026 j_s(t)$ of Fe|Pt and $0.51 j_s(t)$ of $\gamma$-Fe$_2$O$_3$|Pt. We emphasize that such very good agreement is also observed for the corresponding THz electro-optic signals of Fig. 2(b), as is demonstrated in Fig. 3(c). We confirm explicitly that other signal contributions are negligible: magnetic-dipole radiation due to ultrafast demagnetization of Fe$_3$O$_4$ [Fig. S4] [10,63] and signals due to Fe contamination of Fe$_3$O$_4$ by the nearby Fe reference layer, which would yield a signal similar to that from Fe|Pt [Fig. S5(a)].

To summarize, the spin current in Fe$_3$O$_4$|Pt can be very well represented by a superposition of spin currents in two very different samples comprising insulating and conducting magnetic materials, respectively. This remarkable observation strongly suggests that the spin current in Fe$_3$O$_4$|Pt has contributions from both the PSE, i.e., through spin-polarized electrons, [see (1) in Fig. 1(a)] and the SSE, i.e., through spin torque and magnons [see (2) in Fig. 1(a)].

**Physical interpretation for Fe$_3$O$_4$|Pt**

We suggest the following scenario to explain the coexistence of SSE and PSE in Fe$_3$O$_4$.

*SSE.*--Regarding the SSE, we note that the pump excites mainly Pt and, thus, establishes a temperature difference between Pt electrons and Fe$_3$O$_4$ magnons, leading to the SSE spin current across the Fe$_3$O$_4$/Pt interface [Fig. 1(a), (2)]. From the measured spin-current amplitudes [Fig. 3(b)], we infer that the spin-mixing conductance of the Fe$_3$O$_4$/Pt interface is a factor 7.3 larger than that of YIG/Pt [see Table B1], in excellent agreement with literature [62,64,65]. The sign of the current agrees with that of YIG|Pt, suggesting the SSE in Fe$_3$O$_4$ is dominated by the A and B spin-sublattices, whose total magnetization is parallel to the external magnetic field, whereas the e-sublattice is oppositely magnetized.

*PSE.*--Regarding the PSE, we note that the pump also excites the hopping electrons of Fe$_3$O$_4$, either directly by optical absorption in Fe$_3$O$_4$ or by ultrafast heat transport from Pt to the interfacial Fe$_3$O$_4$ regions. Because magnetic order of the e-sublattice is understood to decrease with increasing temperature [44-47], the spin voltage of the e-sublattice electrons changes upon arrival of the pump [Fig. 1(b),(c)] and, thus, triggers spin transfer to the crystal lattice and/or the adjacent Pt layer [Fig. 1(a), (1)] [10]. Remarkably, as the e-lattice spins are on average aligned antiparallel to the equilibrium magnetization $M_0$ [Fig. 1(b),(c)], the PSE tends to increase the magnitude of the total magnetization in Fe$_3$O$_4$, whereas in Fe, it is decreased. We, thus, interpret the observed opposite sign of the PSE currents in Fe$_3$O$_4$|Pt and Fe|Pt [Fig. 2(b)] as a hallmark of the ultrafast quenching of the residual magnetization of the e-sublattice minority hopping electrons in Fe$_3$O$_4$.

The much smaller amplitude of the PSE current in Fe$_3$O$_4$|Pt than for Fe|Pt can have several reasons. First, the transport of spin-polarized electrons requires charge conservation [66,67] and, thus, an equal back-flow of charges. However, because the Fe$_3$O$_4$ spin polarization at the Fermi level is high (-72%) [51], there are less majority states permitting the backflow of spin-unpolarized electrons from Pt to Fe$_3$O$_4$ [10]. Second, the mobility of the Fe$_3$O$_4$ hopping electrons is likely lower than that of the Fe conduction electrons [44,45]. Third, at room temperature, the magnetization of the e-sublattice is significantly smaller than the total Fe$_3$O$_4$ magnetization [44]. The nonvanishing e-sublattice magnetization inferred here suggests that its ferro-to-paramagnetic transition covers a wide temperature range, possibly because of sample imperfections such as impurities [44].

The relaxation time of the PSE is approximately given by the electron-spin equilibration time $\tau_{es}$. Figure 3(b) suggests that the $\tau_{es}$ values of Fe$_3$O$_4$ and Fe are comparable and of the order of 100 fs. This conclusion is consistent with previous measurements of ultrafast demagnetization of Fe$_3$O$_4$, in which an instantaneous drop of the signal was observed directly after optical excitation [68].

It appears that the PSE dynamics does not significantly perturb the slower SSE dynamics, thereby suggesting that the e-sublattice does not excite magnons of the A, B spins to a sizable extent on time scales below 1 ps. Indeed, in laser-induced magnetization dynamics of $Fe_3O_4$ [68], the instantaneous signal drop was followed by a much larger component with a time constant >1 ns. To summarize, we can consistently assign the PSE current in $Fe_3O_4$ to the demagnetization of the e-sublattice-type minority hopping electrons at the Fermi energy.

**Interface sensitivity**

The relative values of the spin-mixing conductance $g_r^{\uparrow\downarrow}$ as inferred above need to be taken with caution because $g_r^{\uparrow\downarrow}$, $M_{IF}$ and, thus, the SSE are very sensitive to the F/Pt interface properties and, therefore, to the growth conditions of the F|Pt stack [17,69,70]. For instance, as observed for YIG|Pt previously [8], the spin current amplitude may vary by up to a factor of 3 from sample to sample. Different interface properties may also explain the amplitude variations of the THz signals between the various YIG|Pt samples studied here [Fig. S2(b)].

For $Fe_3O_4$|Pt, the SSE contribution is robustly observed for samples with Pt grown at room temperature. However, when the Pt deposition temperature is increased to 720 K, the SSE component disappears [Fig. S5(b)]. We assign this effect to Pt-Fe interdiffusion at the interface, which magnetizes Pt in the vicinity of Fe, as reported previously [71,72].

## IV. CONCLUSION

We study ultrafast spin transport in archetypal F|Pt stacks following femtosecond optical excitation. For the ferri/ferromagnetic layer F, model materials with different degrees of electrical conductivity are chosen. For the magnetic insulators YIG, GIG and $\gamma$-$Fe_2O_3$, our results indicate a universal behavior of the interfacial SSE on ultrafast time scales: The spin current is solely determined by the relaxation dynamics of the electrons in the metal layer, and it is localized close to the F/Pt interface.

Remarkably, in the half-metallic ferrimagnet $Fe_3O_4$ (magnetite), our measurements reveal two spin-current components, which exhibit opposite sign and PSE- and SSE-type dynamics. The SSE component is assigned to magnon excitation of the A, B spin sublattices [see (2) in Fig. 1(a)], whereas the PSE component can consistently be assigned to ultrafast demagnetization of e-sublattice minority-spin hopping electrons [(1) in Fig. 1(a)]. Our results show that measuring heat-driven spin currents faster than their natural sub-picosecond formation time allows one to unambiguously separate SSE and PSE contributions by their distinct ultrafast dynamics.

## ACKNOWLEDGMENTS

The authors acknowledge funding by the German Research Foundation through the collaborative research centers SFB TRR 227 "Ultrafast spin dynamics" (projects A05, B02 and B03), SFB TRR 173 "Spin+X" (projects A01 and B02) and project No. 358671374, the European Union H2020 program through the project CoG TERAMAG/Grant No. 681917, the Spanish Ministry of Science and Innovation through Project No. PID2020-112914RB-I00 and the Czech Science Foundation through project GA CR/Grant No. 21-28876J. P.J.-C. acknowledges the Spanish MECD for support through the FPU program (References No. FPU014/02546 and EST17/00382).

## APPENDIX A: SAMPLE FABRICATION

All investigated F|Pt samples including film thicknesses are summarized in Fig. 2(b) and Fig. S2.

### YIG

Films of F=YIG covering a wide range of thicknesses (3.4 nm-3 µm) are fabricated by three different methods on double-side-polished $Gd_3Ga_5O_{12}$ (GGG) substrates. The film with the smallest thickness of 3.4 nm is epitaxially grown on GGG(111) by pulsed laser deposition (PLD) using a KrF excimer laser. The growth temperature is 1000 K, and the oxygen pressure is 7 Pa. The growth is monitored by *in situ* reflection high-energy electron diffraction (RHEED). A clear RHEED pattern is observed, indicating the film is single-crystalline.

The YIG films with thicknesses 5-120 nm are deposited on GGG(111) using a sputtering system (ULVAC-MPS-4000-HC7) with a base vacuum of $1\times10^{-6}$ Pa. After deposition, annealing at 1070 K in oxygen atmosphere is carried out to further improve the crystal quality and enhance in-plane magnetic anisotropy. Finally, a heavy metal Pt thin film is deposited on all the YIG samples.

The YIG film with the largest thickness of 3 µm is grown by liquid-phase epitaxy (LPE) on a GGG substrate (thickness of 500 µm). For details, we refer to Ref. [73] including the Supplementary Information.

### GIG

A film of F=GIG (thickness of 58 nm) on GGG(001) is fabricated by PLD using a custom-built vacuum chamber (base pressure of $2\times10^{-6}$ Pa) and a KrF excimer laser [74]. The growth is performed at a substrate temperature of 475 °C, an $O_2$ background pressure of 2.6 Pa and a deposition rate of 1.4 nm/min without subsequent annealing. A Pt layer (thickness of 2 nm) is deposited *ex situ* by sputtering deposition. The heteroepitaxial growth of GIG and the absence of impurity phases is confirmed by X-ray diffraction.

### $\gamma$-$Fe_2O_3$, $Fe_3O_4$ and Fe

Layers of F=$\gamma$-$Fe_2O_3$ and $Fe_3O_4$ (thickness of 10 nm) are epitaxially grown on MgO(001) substrates (thickness of 0.5 mm) by PLD [61]. Subsequently, all films are *in-situ* covered by DC sputtering with a thin film of Pt (thickness given in Fig. 2(b) and Fig. S2), followed by a thin film of Fe (thickness of 2.5 nm) on part of each of the F|Pt samples, resulting in F|Pt and F|Pt|Fe stacks on the same substrate [see Fig. S1].

Deposition of adjacent F|Pt and F|Pt|Fe stacks on the same substrate significantly simplifies the THz emission experiments. The two stacks can easily be accessed by lateral shifting into the focus of the femtosecond pump beam, with minimal changes of optical paths [Fig. S1]. The THz signal from the F|Pt|Fe regions of each sample serves as an ideal reference that allows for accurate alignment of the setup and definition of the time-axis origin.

## APPENDIX B: DATA ANALYIS

### Extraction of spin-current dynamics

In the frequency domain, the electrooptic signal $S(\omega)$ is related to the THz field $E(\omega)$ directly behind the sample by multiplication with a transfer function $H(\omega)$ that describes the propagation of the THz wave away from the sample and the response function of the electro-optic detector, i.e., the ZnTe crystal [4,8]. Measurement of $H(\omega)$ using a well-understood reference emitter allows us to retrieve $E(\omega)$ and eventually determine the spin current $j_s(\omega)$ through a generalized Ohm's law [34],

$$E(\omega) = e\, Z(\omega)\theta_{\text{SH}}\lambda_{\text{N}} j_{\text{s}}(\omega). \tag{6}$$

Here, $-e$ is the electron charge, $Z(\omega)$ denotes the sample impedance, $\theta_{\text{SH}} \sim 0.1$ is the spin-Hall angle of Pt, and $\lambda_{\text{N}} = 1\,\text{nm}$ is the relaxation length of the spin current in N=Pt [75].

**Relative spin mixing conductance**

We determine the relative spin mixing conductance using the scaling relation [8]

$$\|S\|_{\max} \propto g_{\text{r}}^{\uparrow\downarrow} M_{\text{IF}} a^3 \frac{A}{d} Z. \tag{7}$$

Here, $\|S\|_{\max}$ is the maximum value of the modulus $|S(t)|$ of the THz emission signal, $g_{\text{r}}^{\uparrow\downarrow}$ is the real part of the spin-mixing conductance, $M_{\text{IF}}$ and $a$ are the saturation magnetization and lattice constant of the individual F layer, respectively. Furthermore, $A$ denotes the total pump absorptance of the F|Pt sample under consideration, $d$ is the Pt-layer thickness, and $Z$ is the THz impedance of the stack. Note that the pump power is assumed to be absorbed in the Pt layer only. All quantities required for the estimation of $g_{\text{r}}^{\uparrow\downarrow}$ are taken from literature or are measured [17]. A summary of the parameters as well as the results for the spin mixing conductance $g_{\text{r}}^{\uparrow\downarrow}$ are shown in Table B1.

| Parameter | F=Fe | YIG | GIG | $\gamma$-Fe$_2$O$_3$ | Fe$_3$O$_4$ | References |
|---|---|---|---|---|---|---|
| **Lattice constant $a$ (nm)** | 0.286 | 1.252 | 1.247 | 0.834 | 0.8396 | [76] [77] [78] [79] [42] |
| **Saturation magnetization $M_{\text{IF}}$ (A m$^2$/kg)** | 222 | 20 | 5 | 400 | 400 | [80] [59] [60] [61] |
| **Pt thickness $d$ (nm)** | 2 − 5 | 5 | 2 | 2.5 | 2.5 | Growth |
| **F\|Pt absorptance $A$** | 55 % | 50 % | 50 % | 50 % | 50 % | Measured |
| **Conductivity $\sigma$ (kS/m)** | ∼ 1000 | ∼ 0.1 | ∼ 0.1 | ∼ 0.35 | ∼ 20 | (M) [40] [41] [42] [43] |
| **Infrared refractive index of substrate** | - | 3.5 | 3.5 | 3.07 | 3.07 | [81] [82] |
| **Rel. impedance $Z(\text{F\|Pt})/Z(\text{YIG\|Pt})$** | - | 1.0 | 2.5 | 0.6 | 0.1 | Calculated |
| **Rel. peak signal $\|S(\text{F\|Pt})\|_{\max}/\|S(\text{YIG\|Pt})\|_{\max}$** | - | 1.0 | 1.6 | 8.0 | 8.8 | Measured |
| **Rel. spin mixing conductance $g_{\text{r}}^{\uparrow\downarrow}(\text{F/Pt})/g_{\text{r}}^{\uparrow\downarrow}(\text{YIG/Pt})$** | - | 1.00 | 1.04 | 1.18 | 7.30 | Inferred from measurements |
| **Spin mixing conductance $g_{\text{r}}^{\uparrow\downarrow}$ ($10^{18}$ m$^{-2}$): previous work** | - | ≈ 1 | - | ∼ 1 | ∼ 6 | [62,64] [65] [83] |

**TABLE B1.** Material parameters of Fe, YIG, GIG, $\gamma$-Fe$_2$O$_3$ and Fe$_3$O$_4$, which are measured, calculated or taken from literature for the determination of the relative spin mixing conductance according to Eq. (7). Note that most of the experimentally acquired parameters are relative values and therefore normalized to the respective values of YIG.

# Supplemental Material

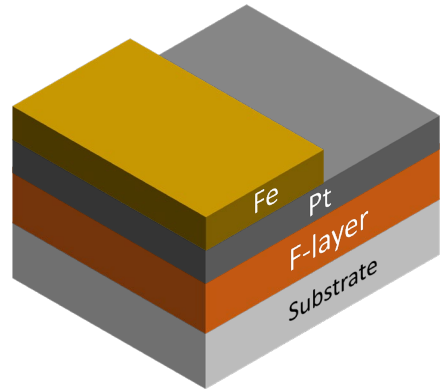

**FIG. S1.** Fe|Pt reference stack. For referencing purposes, an Fe layer is grown on top of part of the F|Pt stack for F=$\gamma$-$Fe_2O_3$, $Fe_3O_4$ and YIG (3μm). The regions with and without Fe can selectively be excited by the laser beam by lateral translation of the sample. For GIG, the stacking order of the Fe|Pt reference layers was reversed, resulting in two regions with thin films of GIG|Fe|Pt and GIG|Pt on the same substrate. The nanometer-thick YIG|Pt samples do not exhibit a reference layer.

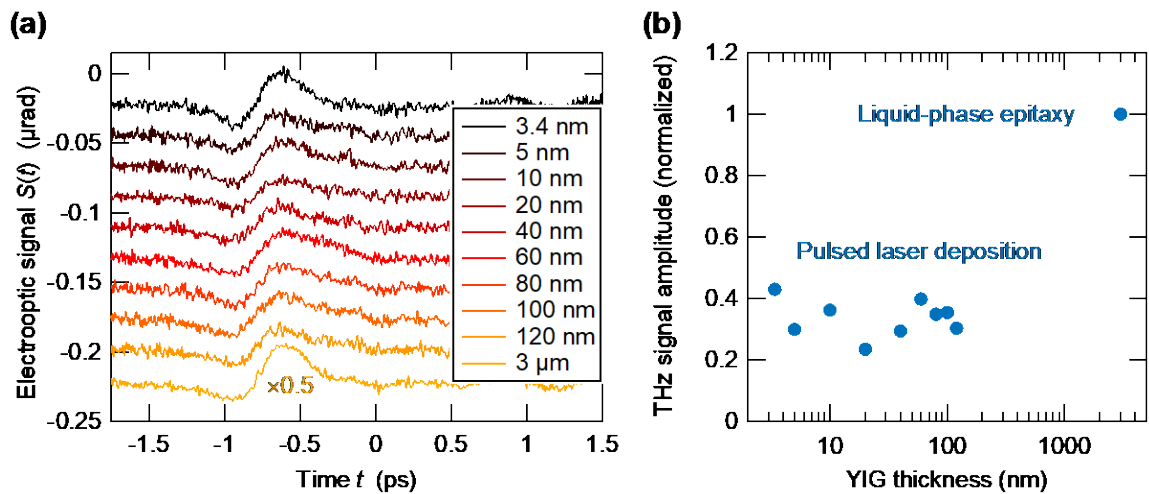

**FIG. S2.** Impact of film thicknesses of the YIG|Pt stacks on THz emission. (a) THz emission signals vs pump-probe delay from YIG($a$)|Pt($b$) with varying YIG thickness ($a = 3.4$ nm and $b = 5$ nm, $a = 5$-$120$ nm and $b = 3$ nm, $a = 3$ μm and $b = 5$ nm). The YIG thin films were grown by pulsed laser deposition, sputtering and liquid-phase epitaxy. (b) Amplitude (root mean square) of the THz signals of YIG($a$)|Pt($b$) vs YIG thickness $a$ normalized to the largest peak signal.

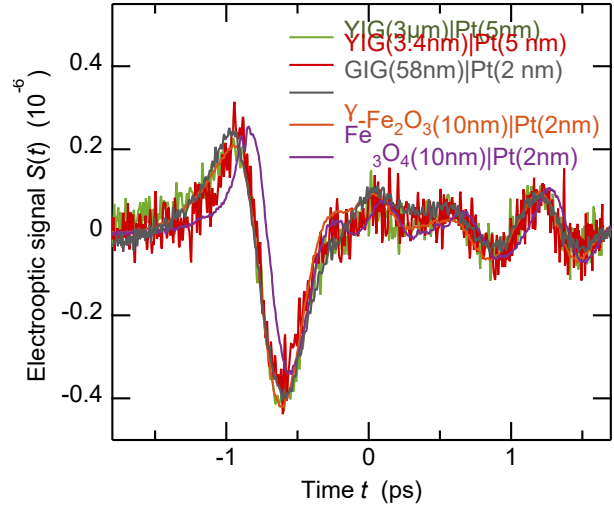

**FIG. S3.** Back-to-back comparison of THz signals. Electrooptic signals of THz pulses emitted from various F|Pt stacks with F=YIG (3 µm and 3.4 nm), GIG, $\gamma$-$Fe_2O_3$ and $Fe_3O_4$ (also see Fig. 2b). All signals are scaled to approximately equal peak amplitude. The time axis has the same origin for all signals and was calibrated by using the signal from the Fe|Pt reference region (see Fig. S1).

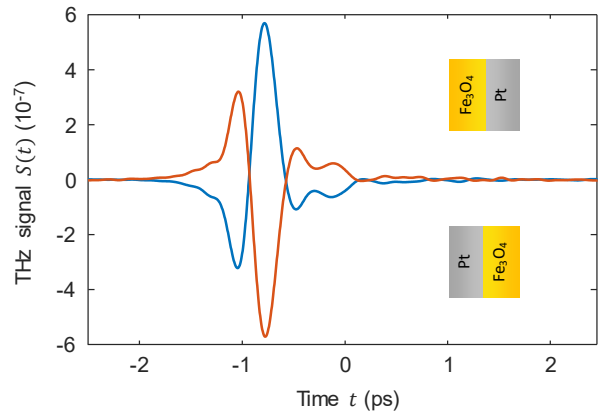

**FIG. S4.** THz electrooptic signals emitted from $Fe_3O_4$|Pt (blue curve) and the Pt|$Fe_3O_4$ sample (orange curve) obtained by 180° turning of $Fe_3O_4$|Pt about an axis parallel to the magnetization $\boldsymbol{M}_0$. The data was corrected for propagation effects of the THz pulse through the substrate and was low-pass filtered with a cut-off frequency of 6 THz. As the two signals are almost perfectly reversed versions of each other, magnetic dipole radiation emitted from $Fe_3O_4$ is not a dominant contribution to the emitted THz signal. Details on the correction procedure can be found in Ref. [10].

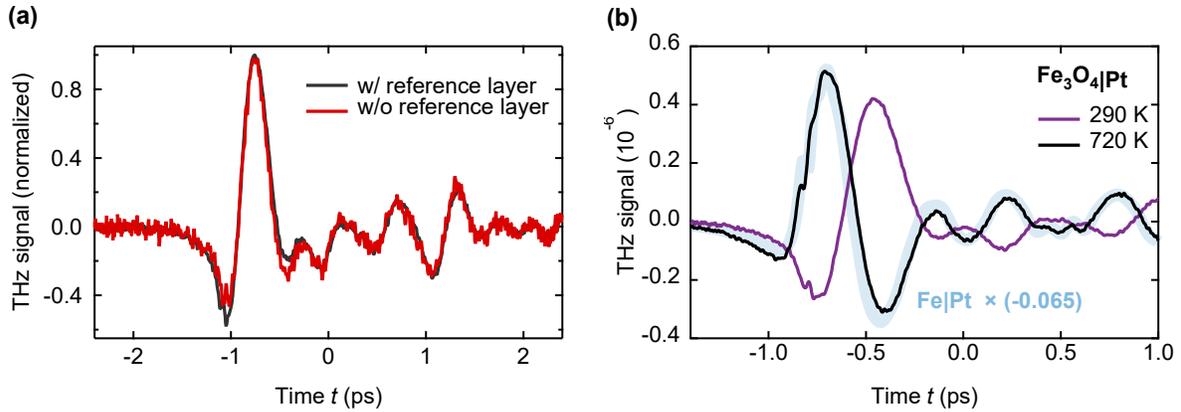

**FIG. S5.** Impact of the Fe reference layer and growth conditions on the THz emission signal from $Fe_3O_4|Pt$. (a) THz signal waveforms from $Fe_3O_4|Pt$ with and without an additional Fe layer in adjacent lateral sample regions (see Fig. S1). As the two signals exhibit almost identical temporal dynamics, we exclude that a sizeable number of Fe atoms is present on top of the nominally Fe-uncovered $Fe_3O_4|Pt$ regions. (b) THz emission signals from $Fe_3O_4|Pt$ stacks for different growth temperatures of the Pt layer: 290 K (violet line) and 720 K (black line). For the latter, the SSE contribution [maximum of the THz electric field at -0.4ps (violet curve) due to slower spin current dynamics] is not observable any more, while the PSE contribution is still present. For comparison, the thick blue line shows the reversed THz signal from Fe|Pt.

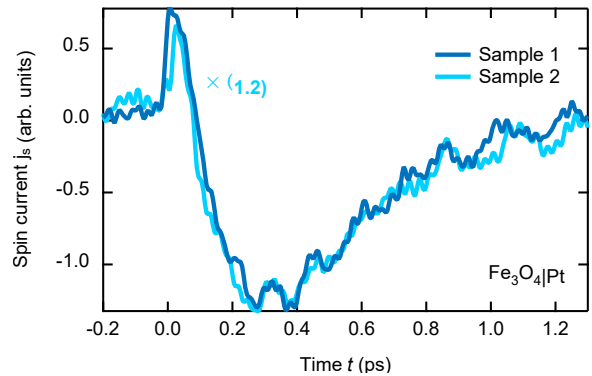

**FIG. S6.** THz spin currents measured for $Fe_3O_4|Pt$ samples that were grown on different days.

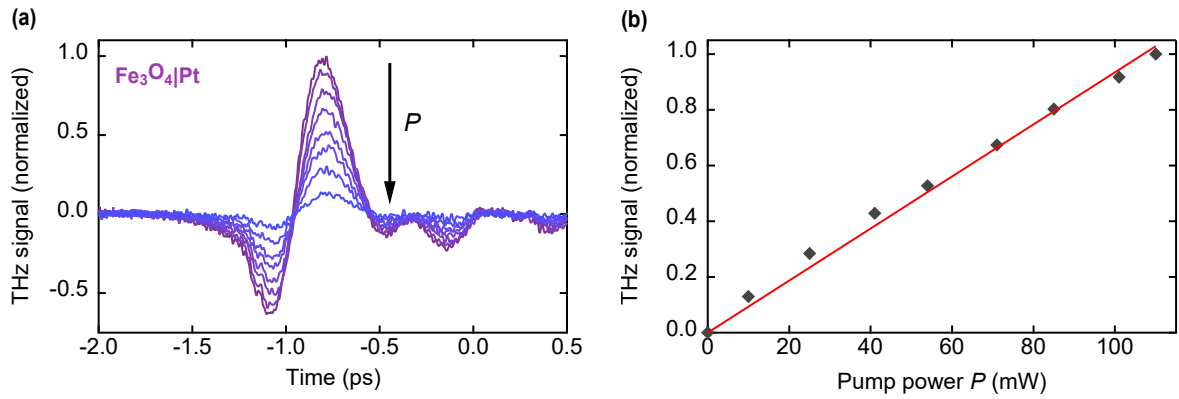

**FIG. S7.** Fluence-dependence of the THz emission signal from $Fe_3O_4|Pt$. (a) THz emission waveforms from $Fe_3O_4|Pt$ for decreasing pump power $P$. The pump power is controlled by a gradient-type neutral-density filter. (b) Root-mean-square amplitude of the waveforms of panel (a) as a function of pump power. The red line shows a linear fit.